\documentclass[a4paper, amsfonts, amssymb, amsmath, reprint, aps, prl, showkeys, nofootinbib, twoside, superscriptaddress]{revtex4-1}
\usepackage[english]{babel}
\usepackage[utf8]{inputenc}
\usepackage[colorinlistoftodos, color=green!40, prependcaption]{todonotes}
\usepackage{multirow} 
\usepackage{amsthm}
\usepackage{xcolor}
\usepackage{graphicx}
\usepackage{times}
\usepackage{comment}

\newcommand{\cG}{\mathcal{G}}

\newcommand{\bk}{\bar{\kappa}}

\usepackage[pdftex, pdftitle={Article}, pdfauthor={Author}]{hyperref} 
\begin{document}
\title{Beyond the Central Limit: Universality of the Gamma Distribution from Pad\'e-Enhanced Large Deviations}

\author{Mario Castro}

\affiliation{Institute for Research in Technology (IIT), Universidad Pontificia Comillas, Grupo Interdisciplinar de Sistemas Complejos (GISC), Madrid, Spain}

\author{Jos\'e A. Cuesta}

\affiliation{Universidad Carlos III de Madrid, Departamento de Matem\'aticas, Grupo Interdisciplinar de Sistemas Complejos (GISC), Legan\'es, Spain}

\affiliation{Instituto de Biocomputaci\'on y F\' {\i}sica de Sistemas Complejos, Universidad de Zaragoza, Zaragoza, Spain}

\date{\today} 

\begin{abstract}
The central limit theorem provides the theoretical foundation for the universality of the normal distribution: under broad conditions, the asymptotic distribution of a sum of independent random variables approaches a Gaussian. Yet, physical systems described by positive random variable---from earthquakes to microbial growth to epidemic spreading---consistently exhibit gamma rather than Gaussian statistics---what leads to field-specific mechanistic explanations that are non robust to small changes in the model details. We show that gamma distributions emerge naturally from large deviation theory when Pad\'e approximants replace polynomial expansions of the derivative of the scaled cumulant generating function, respecting positivity constraints that the central limit theorem violates. Gamma universality thus emerges as the constrained analog of Gaussian universality, providing a mechanism-free explanation for its pervasive appearance across different disciplines.
\end{abstract}


\maketitle

A recurring observation in physics is that complex systems often exhibit statistical regularities that are strikingly independent of microscopic details. These regularities---manifested in universal distributions---serve as guiding motifs for theory-building across disciplines. From the Maxwell-Boltzmann distribution in kinetic theory to the Tracy-Widom law in random matrix theory~\cite{tracy1996orthogonal}, from the Fisher–Tippett–Gnedenko theorem in extreme value statistics~\cite{gnedenko1943distribution} to the ubiquity of power laws in natural and social systems~\cite{bouchaud2000power, stumpf2012critical}, such patterns reflect deep structural principles that transcend specific models.

Among these, the normal distribution stands as the canonical example of universality. The central limit theorem (CLT) provides its theoretical foundation: under broad conditions, the sum of independent random variables converges to a Gaussian, enabling robust predictions in diverse contexts. Yet, this universality has limits. In systems where variables are constrained to be positive, exhibit strong heterogeneity, or are few in number, the Gaussian approximation often fails. In these regimes, usually the gamma distribution emerges—not as a statistical curiosity, but as a persistent empirical regularity.

Indeed, gamma-like distributions appear in the interevent times of earthquakes~\cite{touati_origin_2009,corral_long-term_2004}, in the fracture dynamics of rocks~\cite{davidsen_scaling_2007}, in microbial growth laws~\cite{grilli2020macroecological,camacho2025microbial}, and in the scaling of phenotypic variability in bacteria~\cite{biswas_universality_2024}. They also arise in models of viral entry~\cite{zhang_statistical_2015}, cell division~\cite{brenner_nonequilibrium_2007}, and stochastic exponential growth~\cite{iyer-biswas_universality_2014}. In each case, mechanistic explanations are sought that are system-specific and may attribute causal explanations; however, a unifying theoretical account might clarify this ubiquity but remains elusive.

The limitations of the CLT in these contexts are well understood: convergence may be slow, higher-order corrections remain significant, and positivity constraints impose qualitative changes in behavior. Large deviation theory (LDT) offers a principled extension~\cite{touchette2009large}, but its practical application is often hindered by the difficulty of inverting the rate function to obtain explicit probability densities. Moreover, polynomial approximations of the cumulant generating function (CGF)—which underpin the CLT—can yield invalid densities when higher-order terms are included~\cite{butler2007saddlepoint,daniels:1954}.

In this Letter, we propose a parsimonious and general, yet rigorous, explanation for the emergence of gamma distributions in constrained systems. By applying a Pad\'e approximant~\cite{baker1961application} to the scaled CGF within the LDT framework, we derive gamma-like distributions that outperform the normal approximation even for small system sizes and non-identical variables. This approach naturally respects the positivity constraint and captures essential features of the underlying dynamics. We demonstrate its accuracy across a range of scenarios, including sums of exponential variables with heterogeneous rates, trajectories of Markov chains, and the addition of truncated normal and other non-exponential distributions defined on the positive real axis. Furthermore, we show how the theory can be systematically extended to encompass more complex settings, including convolutions of gamma distributions and non-Markovian dynamics.

Our results suggest that the gamma distribution is not merely a convenient fit, but a universal outcome of constrained aggregation processes. This insight has implications for statistical modeling in physics, biology, and beyond, offering a new lens through which to understand the emergence of regularity in complex systems.

\paragraph{Theory.---}
Consider a sequence of random variables $\mathbf{X}=(X_1,X_2,\dots,X_n)$ with support in $\mathcal{X}$, and let $S_n(\mathbf{X})$ be a function of this sequence (e.g.~the sum of the sequence). The probability density of this variable can be obtained as
\begin{align*}
    p_n(x) &=\int_{\mathcal{X}}\delta\big(S_n(\mathbf{X})-x\big)
    p(\mathbf{X})\,d\mathbf{X} \\
    &=\frac{1}{2\pi i}
    \int_{a-i\infty}^{a+i\infty}d\xi\,e^{-\xi x}
    \left\langle e^{\xi S_n(\mathbf{X})}\right\rangle,
\end{align*}
where we have used the Laplace transform representation of Dirac's delta function
\begin{equation*}
    \delta(u)=\frac{1}{2\pi i}\int_{a-i\infty}^{a+i\infty}e^{\xi u}\,d\xi.
\end{equation*}

Let us denote
\begin{equation}\label{eq:lambda}
    \lambda_n(\xi)\equiv\frac{1}{n}\log\left\langle
    e^{\xi S_n(\mathbf{X})}\right\rangle.
\end{equation}
If $\lambda_n(\xi)$ has a well defined limit when $n\to\infty$, then $S_n$ satisfies a large deviation principle~\cite{touchette2009large}, and the probability density admits the asymptotic representation
\begin{equation}\label{eq:pn_vs_lambda}
    p_n(x)\sim\frac{1}{2\pi i}\int_{a-i\infty}^{a+i\infty}
    e^{n[\lambda_n(\xi)-\xi y]}\,d\xi \quad (n\to\infty),
\end{equation}
where $y\equiv x/n$. The integral in Eq.~\eqref{eq:pn_vs_lambda} can be estimated via the saddle-point method~\cite{daniels:1954}, yielding
\begin{equation}\label{eq:pnasymp}
    p_n(x)\sim\frac{c_n}{\sqrt{\lambda''_n(\xi^*)}}e^{n[\lambda_n(\xi^*)-\xi^*y]}
    \quad (n\to\infty),
\end{equation}
where $\xi^*$ solves the saddle-point equation
\begin{equation}\label{eq:saddle}
    n\lambda'_n(\xi^*)=x,
\end{equation}
and $c_n$ is a normalization constant.

The classical central limit theorem (CLT) amounts to making the linear approximation
\begin{equation}\label{eq:Taylor}
    n\lambda'_n(\xi)\approx\mu_n+\sigma^2_n\xi,
\end{equation}
where $\mu_n$ and $\sigma_n^2$ are the mean and variance of $S_n$. Within this approximation and using Eq.~\eqref{eq:saddle},
\begin{equation*}
    n\lambda_n(\xi)=\mu_n\xi+\frac{\sigma^2_n}{2}\xi^2, \qquad
    \xi^*=\frac{x-\mu_n}{\sigma_n^2},
\end{equation*}
whereby $n\lambda_n(\xi^*)-\xi^*x=-(x-\mu_n)^2/2\sigma_n^2$. As $\lambda_n''(\xi^*)$ is a constant, Eq.~\eqref{eq:pnasymp} becomes the normal distribution.

Inspired by Baker's work on Pad\'e approximants in Statistical Physics~\cite{baker1961application}, we propose using rational approximations to $\lambda'_n(\xi)$. To lowest order, the $[0/1]$-Pad\'e approximant becomes
\begin{equation}\label{eq:cgfslope}
    n\lambda_n'(\xi)\approx\frac{\mu_n}{1-\sigma_n^2\xi/\mu_n}.
\end{equation}
This approximation offers an advantage over \eqref{eq:Taylor}: it is positive in the whole interval of validity of the Pad\'e ($\xi<\mu_n/\sigma_n^2$)---hence \eqref{eq:saddle} yields $x>0$. Thus, for distributions with support $x>0$, \eqref{eq:cgfslope} provides the lowest-order rational approximation that preserves this constrained support.

Now, from \eqref{eq:cgfslope},
\begin{equation*}
    n\lambda_n(\xi)=-\frac{\mu_n^2}{\sigma^2_n}
    \log\left(1-\frac{\sigma_n^2}{\mu_n}\xi\right), \quad
    \xi^*=\frac{\mu_n}{\sigma_n^2}\left(1-\frac{\mu_n}{x}\right),
\end{equation*}
and therefore
\begin{equation*}
    n\lambda_n(\xi^*)-\xi^*x=\frac{\mu_n^2}{\sigma^2_n}
    \log\left(\frac{x}{\mu_n}\right)-\frac{\mu_n}{\sigma_n^2}x+\text{const.}
\end{equation*}
As $\lambda''_n(\xi^*)\propto x^2$, the resulting approximation is the gamma distribution
\begin{equation}\label{eq:gamma}
    p^{(G)}_n(x)\sim c_n\left(\frac{x}{\mu_n}\right)^{\alpha_n-1}
    e^{-\alpha_n x/\mu_n}, \quad \alpha_n\equiv
    \frac{\mu_n^2}{\sigma_n^2}.
\end{equation}
This result is exact for sums of independent and identically distributed (i.i.d.) exponential variables---in which case, the resulting gamma distribution is also known as the Erlang distribution.

The fact that \eqref{eq:gamma} is a direct consequence of \eqref{eq:cgfslope} reveals that gamma's universality is the natural analog of Gaussian's universality for distributions with domain $x>0$.

\paragraph{Numerical tests.---} To quantify the accuracy of Eq.~\eqref{eq:gamma}, we compute the Kullback-Leibler divergence (KLD)
\begin{equation}\label{eq:KL}
    D_{\text{KL}}(p_n \,\|\, p^{\text{app}}_n) = \int_{-\infty}^{\infty} p_n(x)
    \log\left(\frac{p_n(x)}{p_n^{\text{app}}(x)}\right)\,dx
\end{equation}
between the exact distribution $p_n$ and any approximation $p^{\text{app}}_n$. This divergence vanishes if, and only if, the approximation is exact. 

We explore three scenarios: (i) non-identical, independent exponential distributions; (ii) truncated normal distributions; and (iii) non-identical generalized gamma distributions (such as those arising in microbiome ecology~\cite{camacho2025microbial}).

\paragraph{(i) Non-identical, independent exponential distributions:}
Let $X_i$, $i=1\ldots n$, be independent random variables with exponential distributions
\begin{equation}
    p(X_i|\lambda_i)=\lambda_ie^{-\lambda_iX_i},
\end{equation}
where the rates $\lambda_i$ are drawn from different \emph{hyper-distributions} $q(\lambda)$. We want to approximate the distribution of $S_n=X_1+\cdots+X_n$ by either a normal or a gamma distribution with the exact mean and variance.

\begin{figure*}
    \begin{tabular}{lll}
     a) & b) & c)\\ 
         \includegraphics[width=0.33\textwidth]{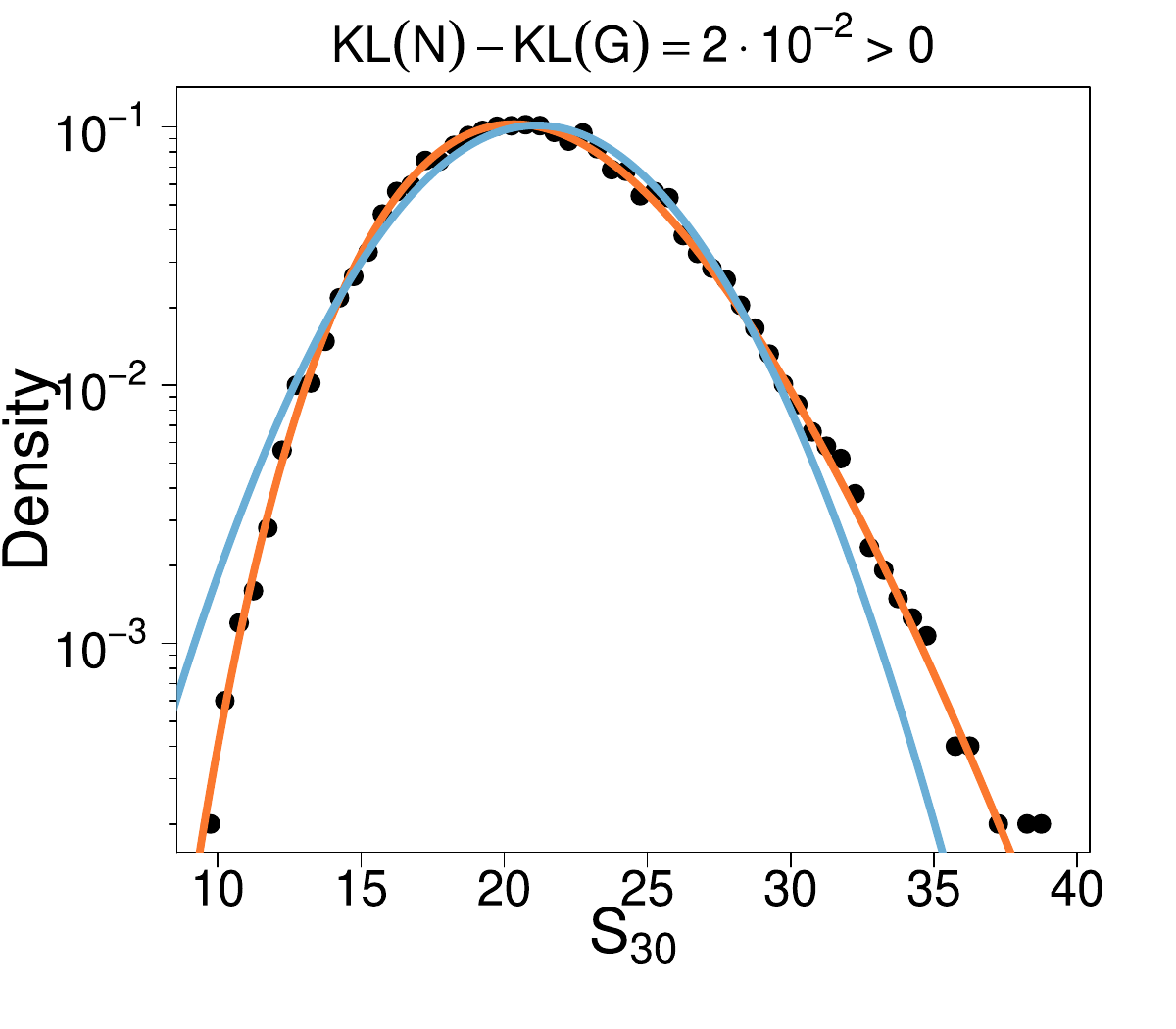}  &
         \includegraphics[width=0.33\textwidth]{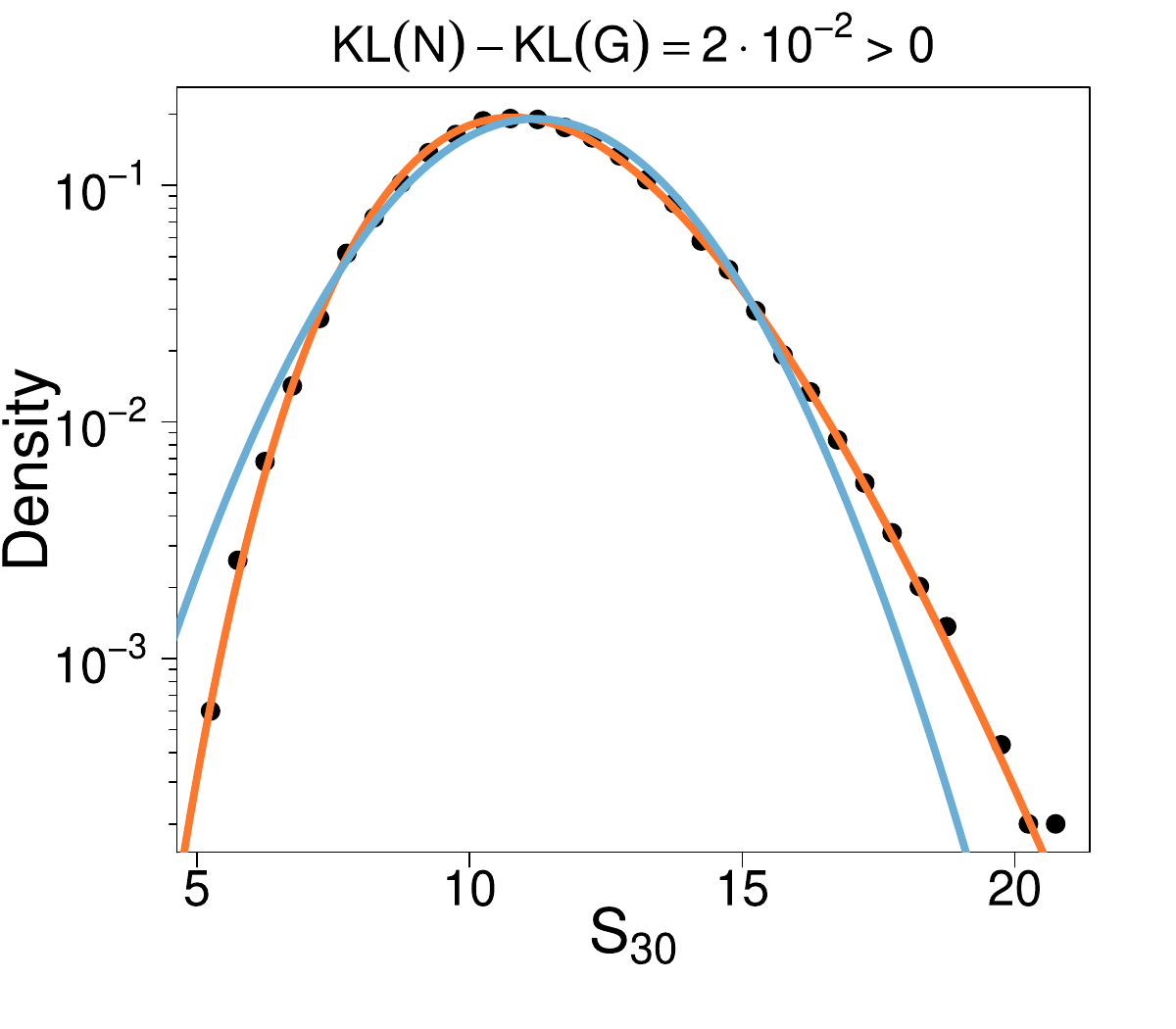}     &
\includegraphics[width=0.33\textwidth]{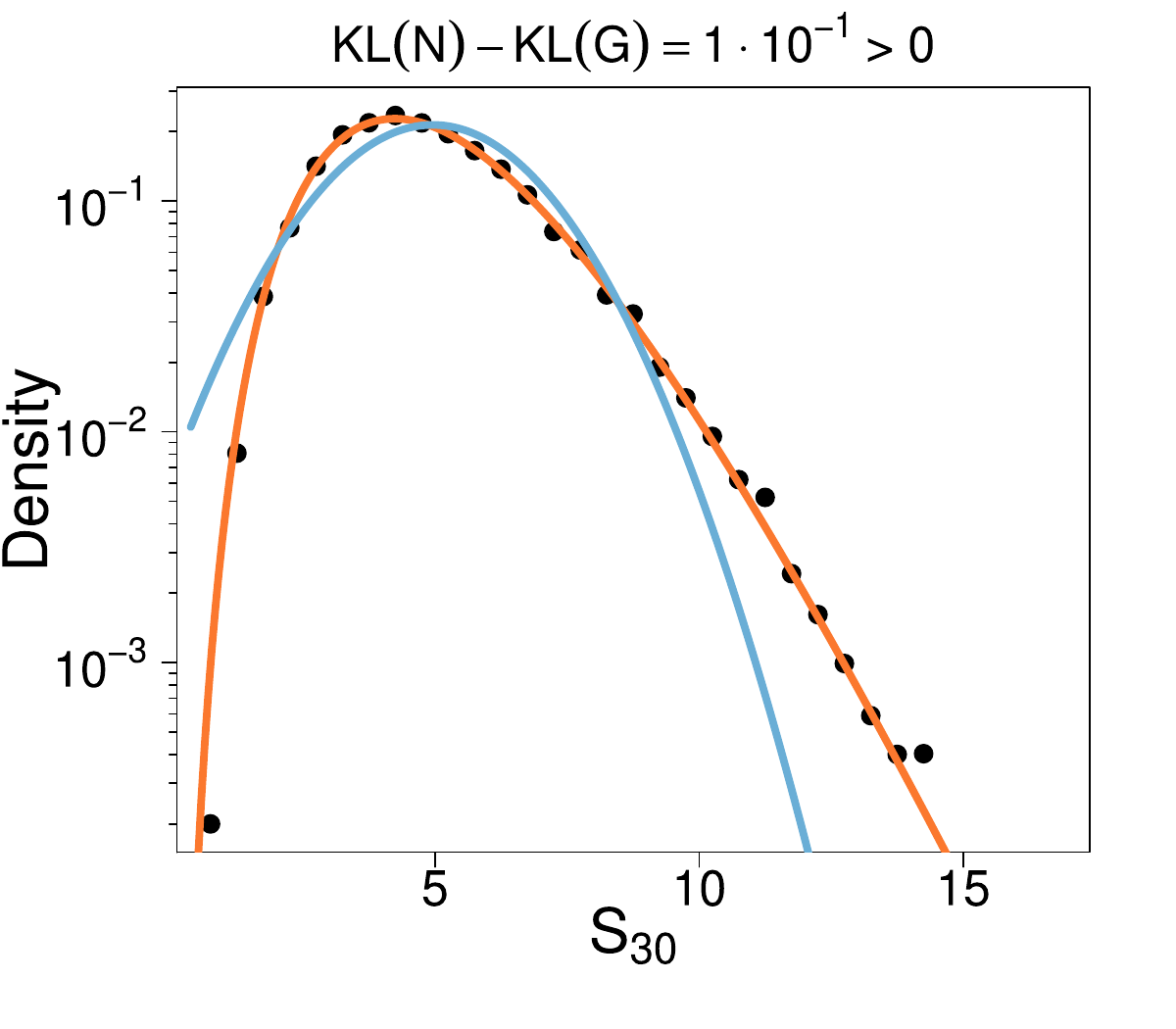} \\    
              d) & e) & f)\\ 
              \includegraphics[width=0.33\textwidth]{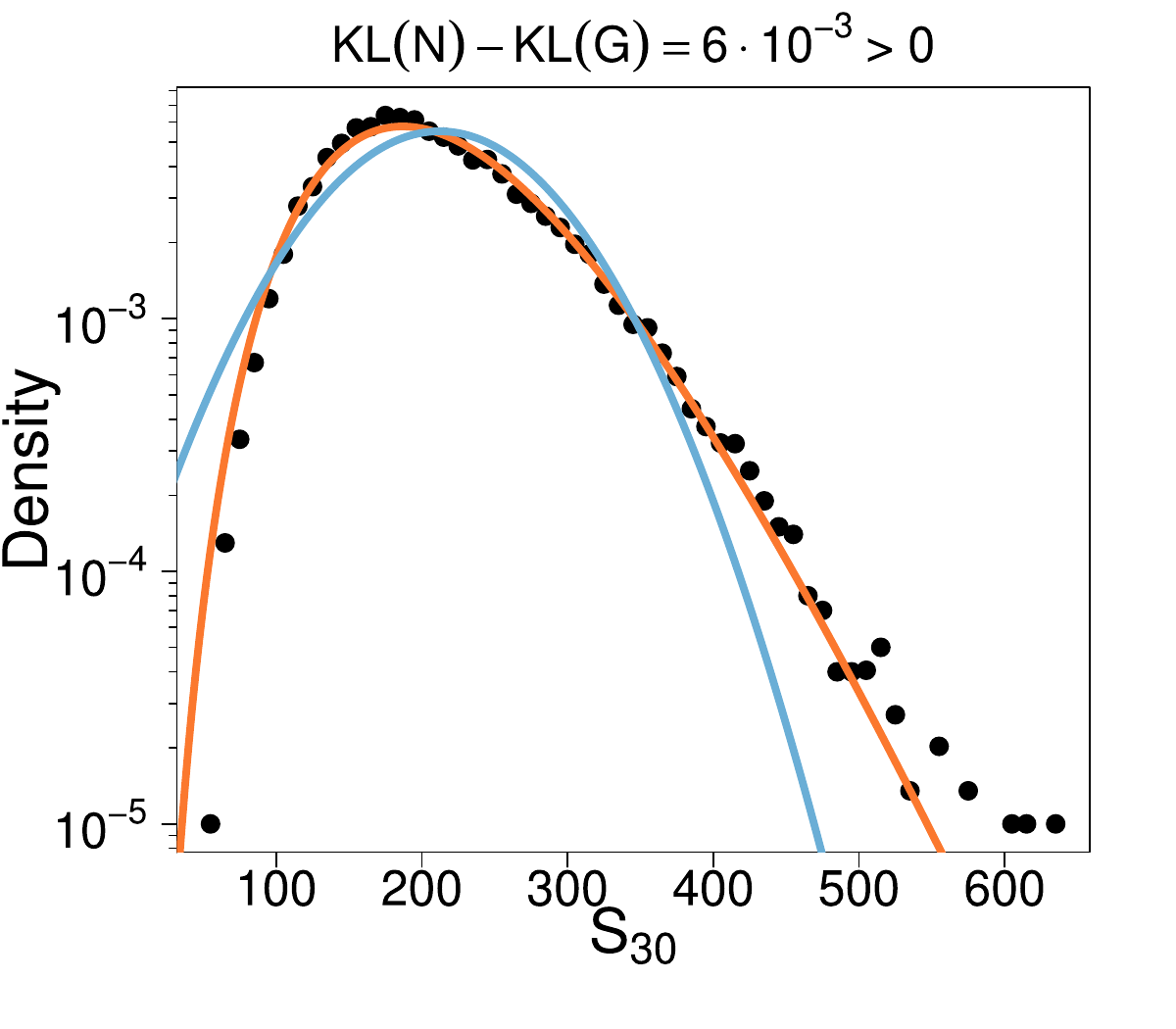}  & 
         \includegraphics[width=0.33\textwidth]{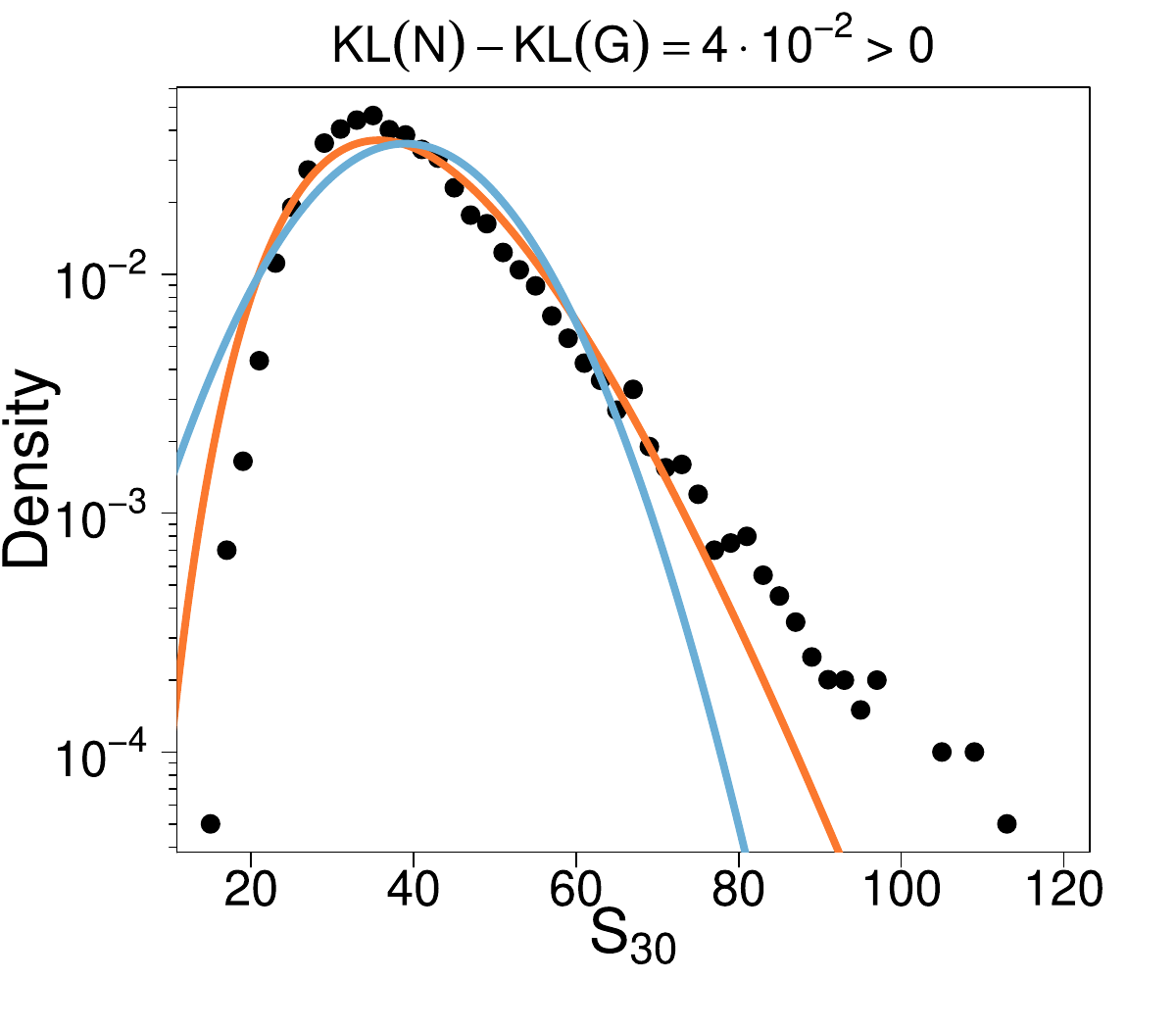}  &    
         \includegraphics[width=0.3\textwidth]{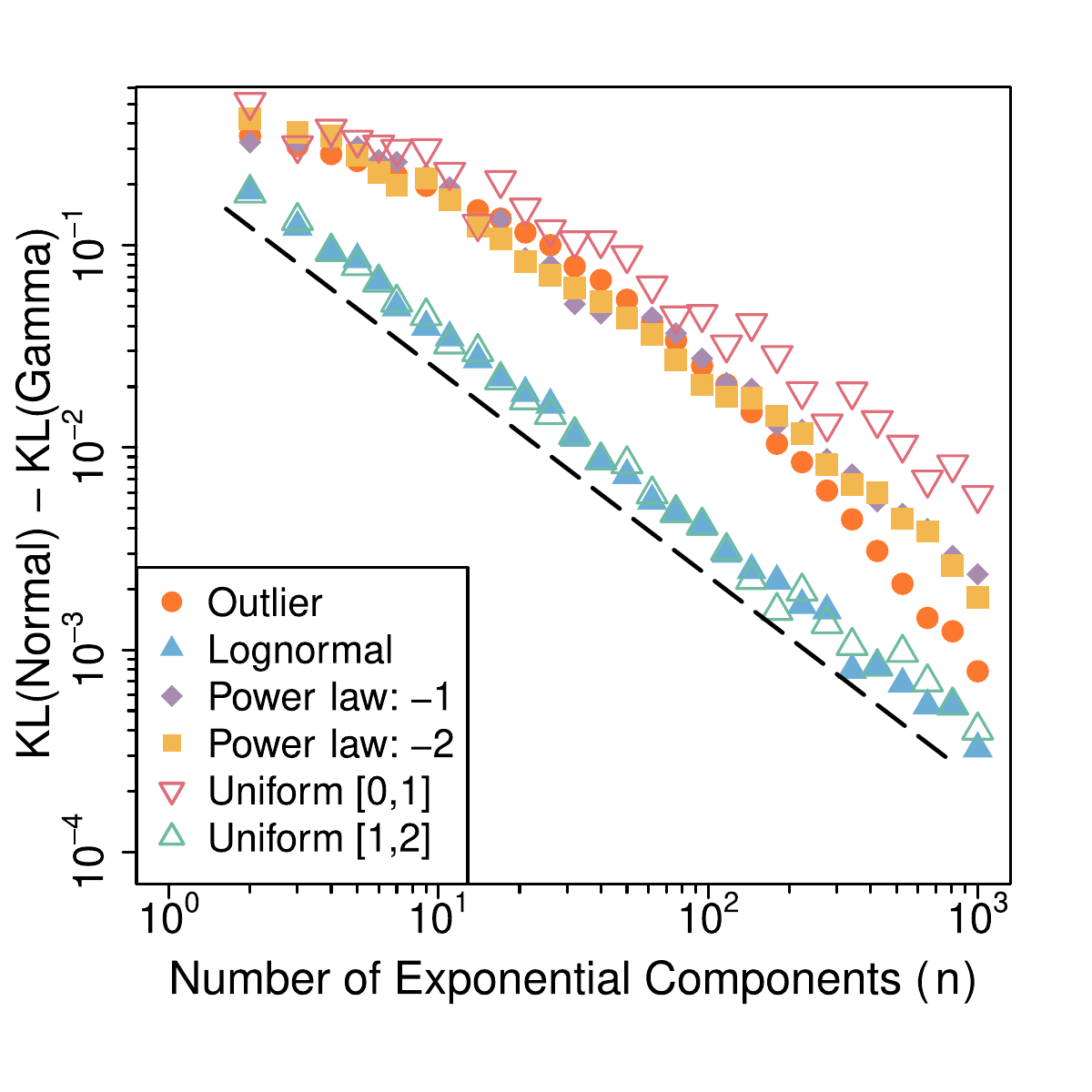}            \\
    \end{tabular}
    \caption{\label{fig:comparison30} Empirical distributions of the sum $S_n$ of $n=30$ exponential variables with rates drawn from various hyper-distributions $q(\lambda)$.  (a) Uniform[1,2]; (b) Log-normal(1,2); (c) power-law $q(\lambda)\sim\lambda^{-1}$; (d) Uniform[0,1]; (e) One outlier rate (see text); (f) KL divergence difference between normal and gamma. The dashed straight line scales as $\sim 1/n$. The gamma approximation (orange line) consistently outperforms the normal approximation (blue line). In the title, the KL divergence difference is always positive (the gamma is better). The cases in the bottom row show some discrepancies in the tails, due to the presence of some rate(s) $\lambda_i\simeq 0$ (see text).
    }
\end{figure*}

Figure~\ref{fig:comparison30} shows the empirical distributions of $S_{30}$ for several choices of $q(\lambda)$. The gamma approximation consistently outperforms the normal. We can justify this success by using an estimate of the KLD obtained from the Edgeworth expansion~\cite{jondeau2001gram} (see section S1 of the Supplemental Material~\cite{SM} and Refs [1-3] therein). If $p_n^{\text{app}}$ is either a normal or a gamma distribution, the leading term of the expansion is
\begin{equation}
    D_{\text{KL}}(p_n \,\|\, p^{\text{app}}_n)\approx
    \frac{1}{12}\big(\bk_{3}-\bk_3^{\text{app}}\big)^2,
    \label{eq:kappa3}
\end{equation}
where $\bk_k\equiv\kappa_{k,n}/\sigma_n^k$ is the normalized $k$th order cumulant, and the superscript `app' refers to the approximating distribution.
Using the Cauchy-Schwarz inequality, we can prove \cite[Sec.~S1]{SM} that for a sum of exponential random variables $\big(\bk_3-\bk_3^{\mathcal{G}}\big)^2<\bk_3^2$, so for sufficiently large $n$, $D_{\text{KL}}(p_n \,\|\, \phi_{\mathcal{G}})<D_{\text{KL}}(p_n \,\|\, \phi_{\mathcal{N}})$. In other words, \emph{the gamma approximation outperforms the central limit approximation} precisely in the limit where the latter is supposed to dominate.

Returning to Fig.~\ref{fig:comparison30}, we can see that, in general, the gamma approximation captures the empirical distribution over the whole range of parameters. The only discrepancies that one can spot occur at the extreme tails, only for some particular cases: those where some rates are too close to $0$ ($[0,1]$-uniform distribution); when they span several orders of magnitude (harmonic case, with rates $\lambda_i=i$, for $i=1, \ldots, n$); or when one rate is $10$ times lower (outlier) than the rest---which are identical.

In order to gain insight into the latter case, we can analyze the hyper-distribution $q(\lambda)=(1-1/n)\delta(\lambda-1)+(1/n)\delta(\lambda-\alpha)$. Then, if $n$ is large enough \cite[Sec.~S2]{SM},
$$
D_{\mathrm{KL}}\!\bigl(p_n\,\Vert\,\phi_{\mathcal G}\bigr)
\approx\frac{(n-1)^2\alpha^{-2}\big(1-\alpha^{-1}\big)^4}
{3\big(n-1+\alpha^{-2}\big)^3\big(n-1+\alpha^{-1}\big)^2}.
\label{eq:KLalpha}
$$
For large $\alpha$ this scales as $D_{\mathrm{KL}}\!\bigl(p_n\,\|\, \phi_{\mathcal G}\bigr)\sim 1/3\alpha^2n^3$, whereas for small $\alpha$ it scales as $\sim\alpha^2n^2/3$. The former scaling justifies why outliers with a higher rate are well captured by the gamma approximation, whereas low-rate outliers spoil the approximation at the tails \cite[Fig.~S1]{SM}.

\begin{figure}[!htp]
    \centering
    \begin{tabular}{ll}
    a) & b) \\
      \includegraphics[width=.24\textwidth]{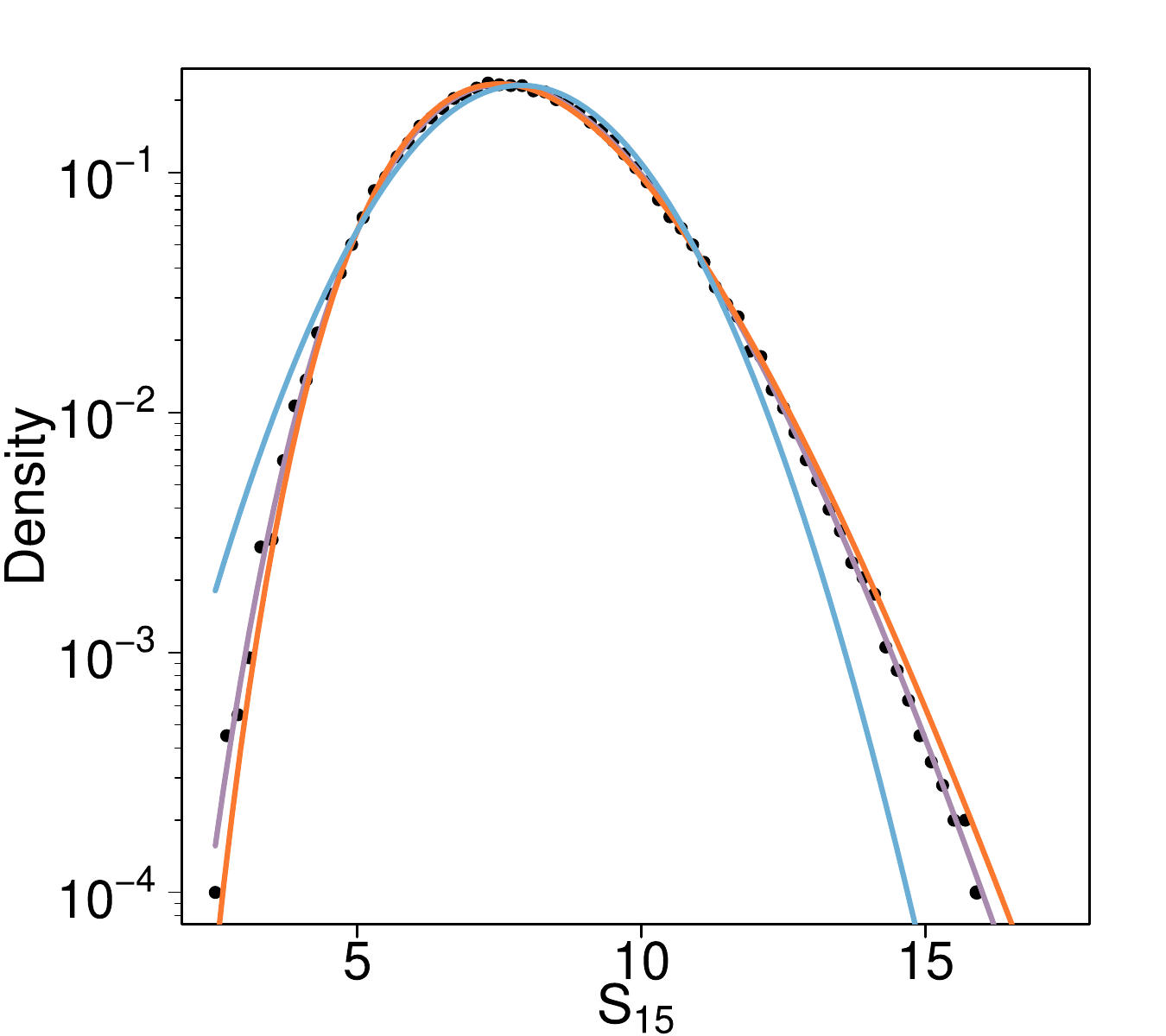}
         &  \includegraphics[width=.24\textwidth]{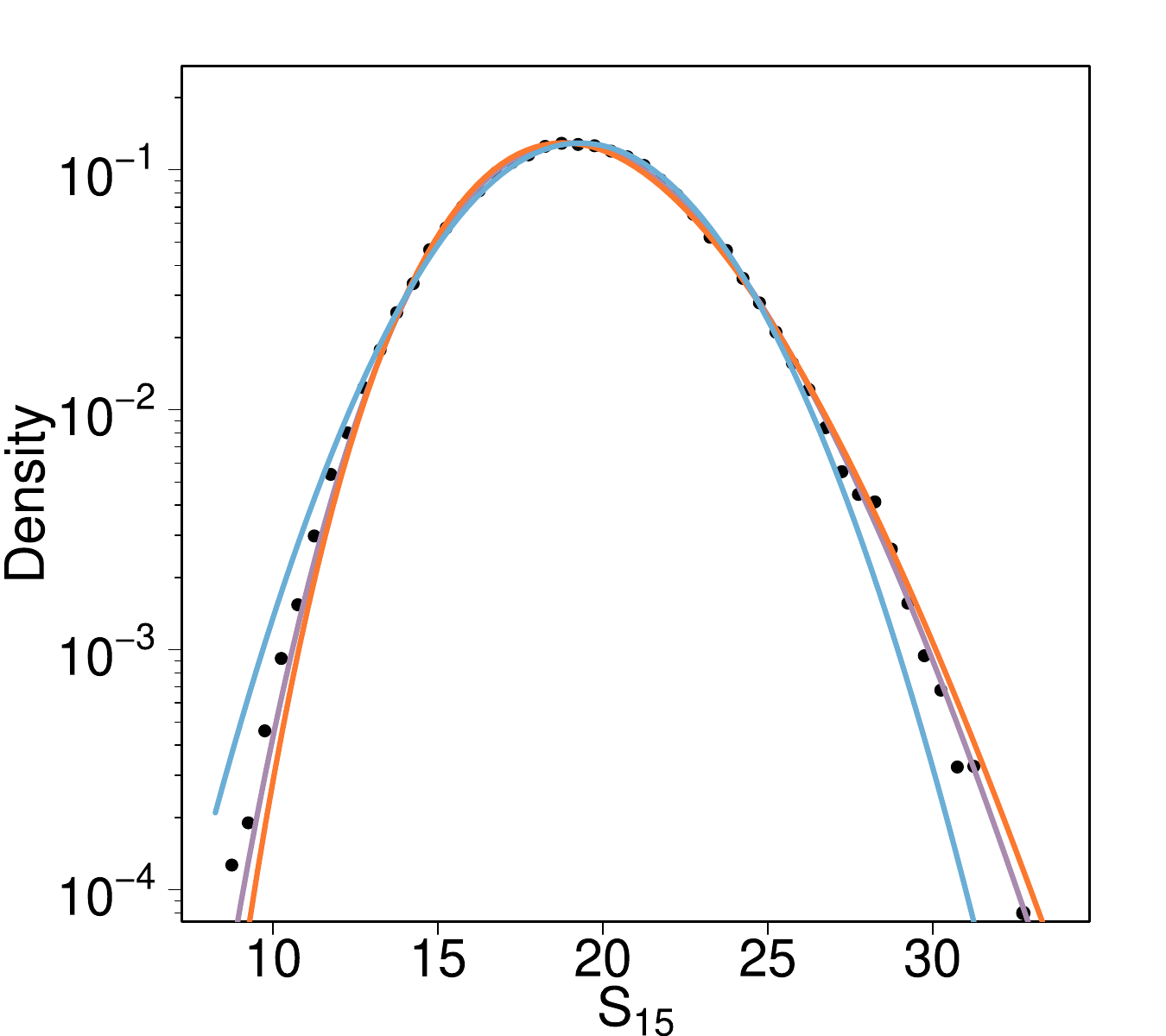}\\
         c) & d)\\
      \includegraphics[width=.24\textwidth]{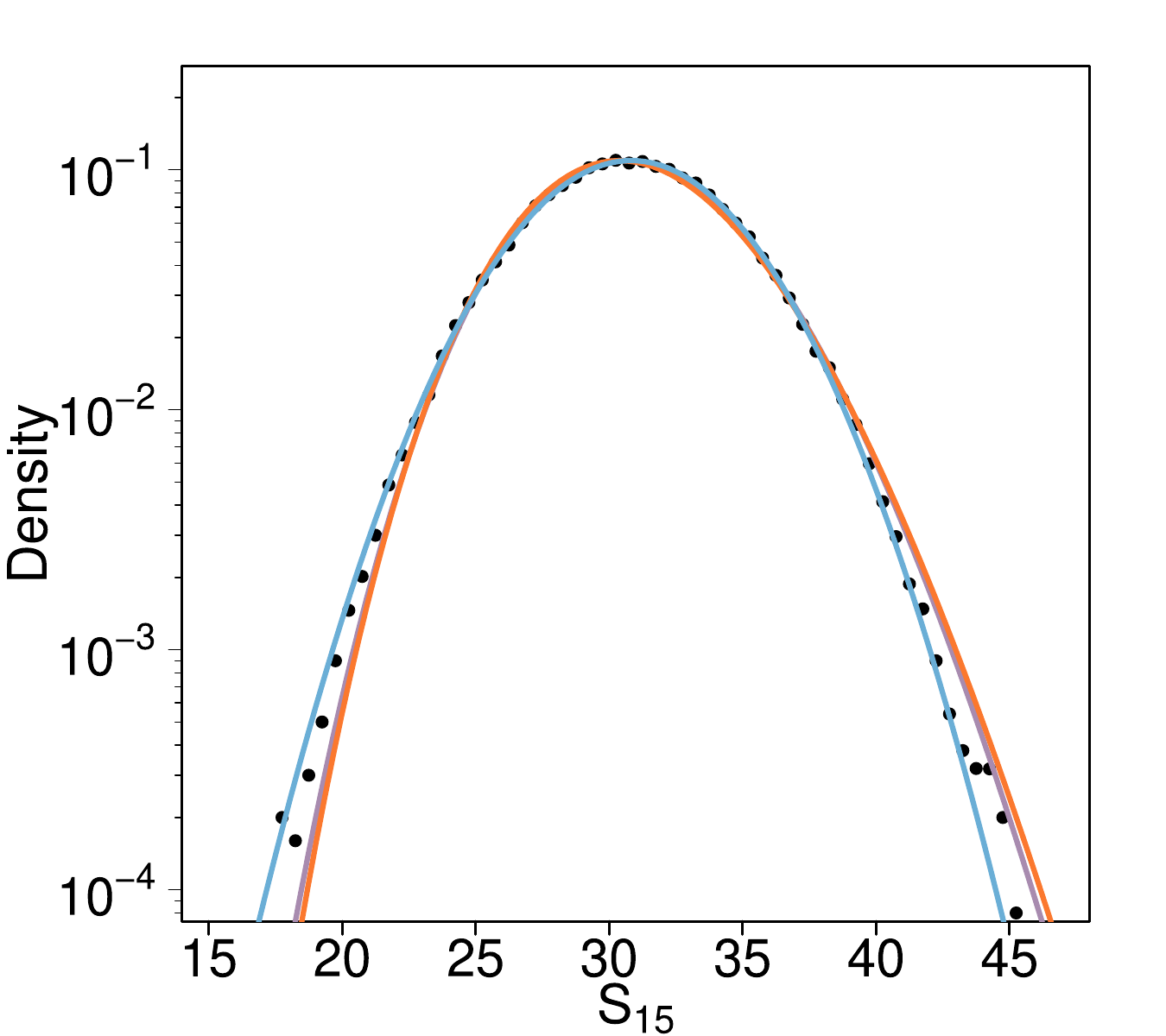} & 
        \includegraphics[width=.24\textwidth]{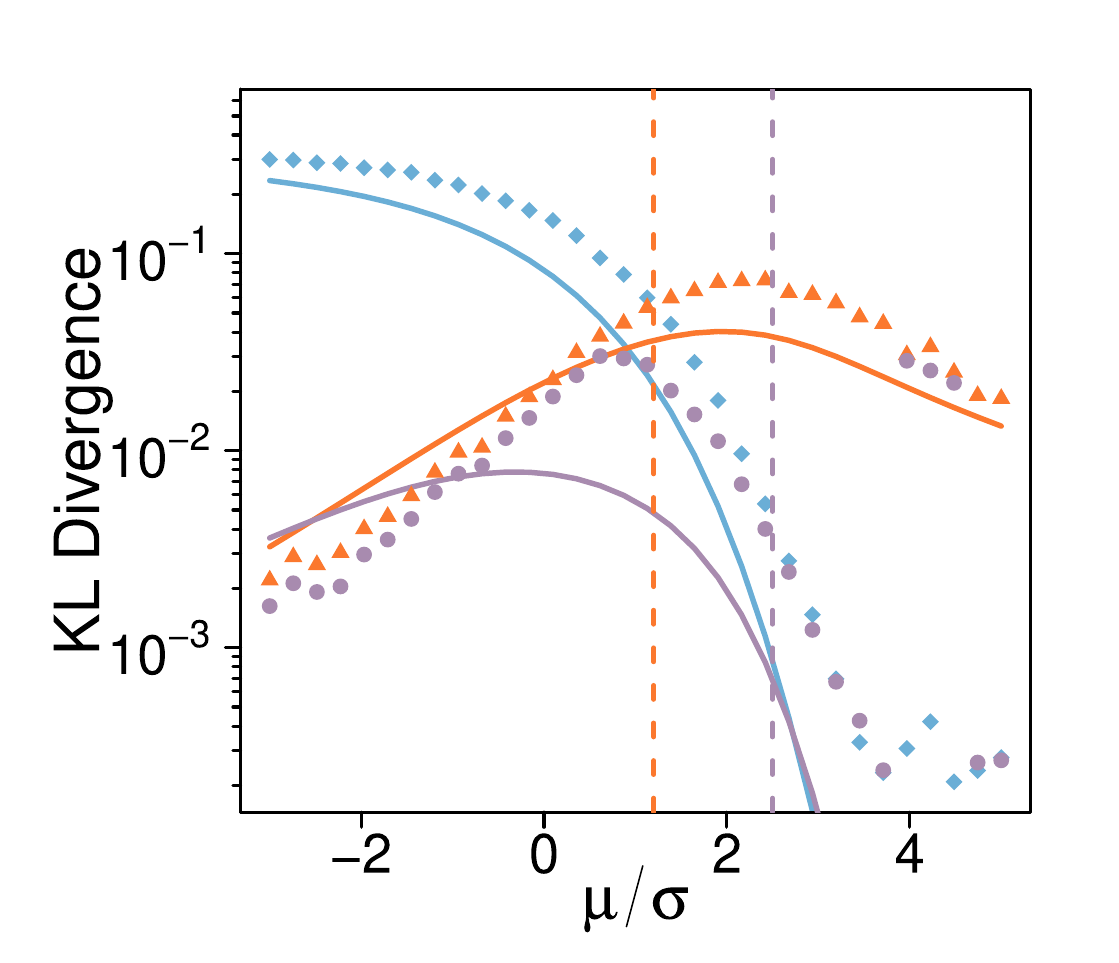}
    \end{tabular}
    \caption{\label{fig:truncnorm}Empirical distribution of $S_{15}$ and the corresponding normal (blue diamonds), gamma (orange triangles), and shifted-gamma (purple circles) approximations for the sum of variables drawn from a truncated normal distribution with different values of $\mu$. a) $\mu/\sigma=-1$; b) $\mu/\sigma=1$; c) $\mu/\sigma=2$; d) Numerical KLD (symbols) and approximation Eq.~\eqref{eq:kappa3}, for the normal and the gamma, and Eq.~\eqref{eq:kappa4} for the shifted gamma. Note how at $\mu/\sigma\sim 1$ (orange dashed line for the gamma, and purple dashed for the shifted gamma), both KLD cross, meaning that above that, the normal approximates better the empirical distribution, and below, the gamma is. Also note how Eqs.~\eqref{eq:kappa3}-\eqref{eq:kappa4} capture the transition qualitatively.  In the case of the shifted gamma, the approximation extends up to $\mu/\sigma\simeq 2$ as the third cumulant can also be matched (see Supplemental Material, Table~S2).}
\end{figure}

\paragraph{(ii) Truncated normal distributions:}
Although the Pad\'e expansion \eqref{eq:cgfslope} does not assume that the elements of $S_n$ must be exponentially distributed, one may wonder whether the success of the gamma is somehow linked to the exponential decay of the empirical distribution. Our next scenarios show that the improvement gained by using Pad\'e approximants goes beyond the sum of exponential random variables. For instance, consider $X_i$ to be drawn from a positively truncated normal distribution $p(X_i)\propto e^{-(X_i-\mu)^2/\sigma^2}\Theta(X_i)$. In Fig.~\ref{fig:truncnorm}, we show that if the left tail of the normal distribution is heavily truncated, the gamma distribution better describes again the distribution of $S_n$. We quantify this by numerically computing the KLD, alongside Edgeworth approximations that qualitatively capture its shape \cite[Sec.~S3]{SM}. As shown in Fig.~\ref{fig:truncnorm}d), only when $\mu\gtrsim\sigma$ does the normal outperform the gamma.

\paragraph{(iii) Non-identical generalized gamma distributions:}
This last scenario highlights the practical implications of our work in an ongoing debate in ecology. Recent work has proposed the existence of universal macroecological laws, including the gamma distribution of abundance fluctuations~\cite{grilli2020macroecological}. The claim is that this universality provides significant insight into the microscopic laws governing ecosystem dynamics. However, the mechanistic models proposed to justify the appearance of such a distribution \cite{grilli2020macroecological, camacho2024sparse, george2023universal} are sensitive to specific choices of certain terms. As a matter of fact, disaggregating by ecological niches, other distributions seem to fit the data as accurately, such as the generalized gamma~\cite{camacho2025microbial}
\begin{equation}
f(x)=\frac{\lambda^{\theta k}\theta}{ \Gamma(k)} x^{\theta k-1} e^{-(\lambda x )^\theta} \quad (x\ge 0).
    \label{eq:gengamma}
\end{equation}
Note that for $\theta=1$ the distribution is a gamma and for $\theta=2$ and $k=1/2$ a truncated normal with $\mu=0$.

\begin{figure}[!htp]
    \centering
    \begin{tabular}{ll}
    a) & b)\\
         \includegraphics[width=.25\textwidth]{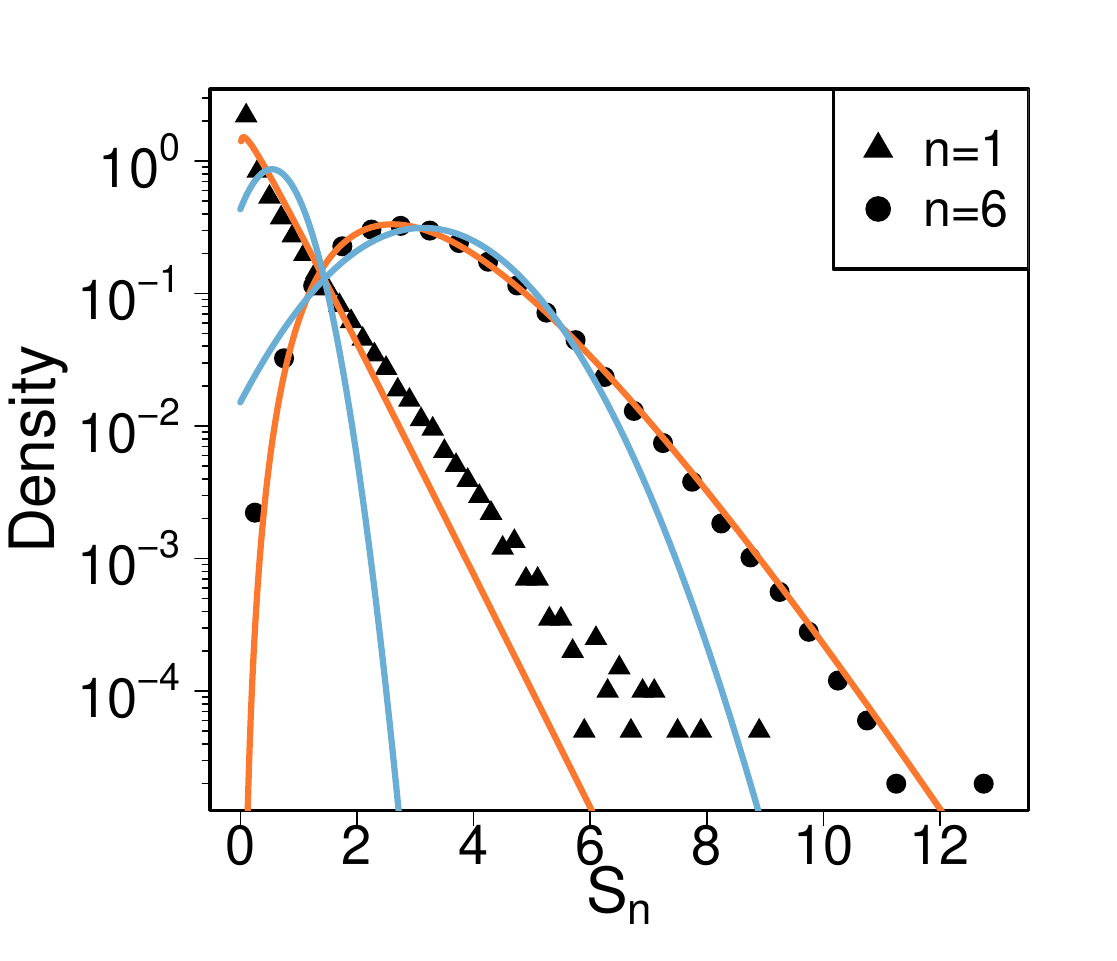} &
         \includegraphics[width=.25\textwidth]{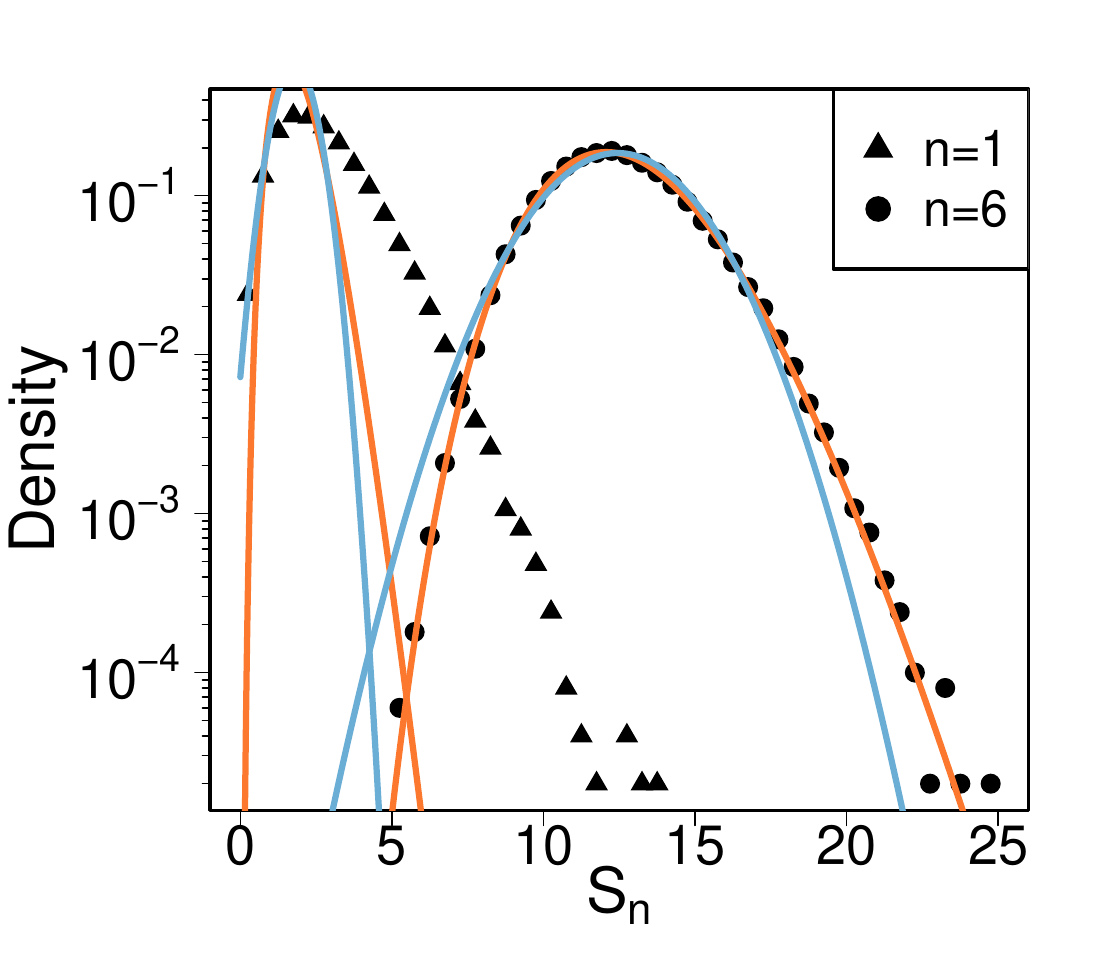} \\
    \end{tabular}
        
    \caption{Gamma (orange) and normal (blue) approximations of the empirical histogram for a single (triangles) generalized gamma ($\theta=1.75$) and the sum (circles) of generalized gamma random variables from $n=6$ ecological niches~\cite{camacho2025microbial} with parameters $\lambda=1$, $\theta_1,\ldots,\theta_6=1.49, 1.54, 1.11, 1.60, 1.75, 1.42$, respectively, and a) $k=0.50$ and b)  $k=3$. }
    \label{fig:gengamma}
\end{figure}

In our final numerical experiment, we consider $n=6$ different ecological subpopulations, each following a generalized gamma distribution with parameters $\theta$ taken from the maximum posterior distribution of Ref.~\cite{camacho2025microbial}. As $\lambda$ simply sets the scale of the distribution, we fix $\lambda=1$.  To test the generality of the approximation, we vary the shape parameter $k$. As shown in Fig.~\ref{fig:gengamma}, in spite that the distribution of each random variable is poorly captured by a gamma, it nevertheless becomes accurate by aggregating just $n=6$ of them (see also \cite{SM} for further cases). Considering that the tails of the generalized gammas are not exponential, this result is remarkable.

\paragraph{Beyond the gamma approximation.---}
One advantage of using Pad\'e approximants in \eqref{eq:cgfslope} is that, whereas improving the Taylor expansion may lead to negative probabilities, Pad\'es ensure positivity. For instance, if we chose a $[1/1]$-Pad\'e,
\begin{equation}
n\lambda_n'(\xi)\approx \mu_n+\frac{\sigma_n^2\xi}{1-
\kappa_{3,n}\xi/2\sigma_n^2},
\end{equation}
we would obtain shifted gamma distribution. This distribution often improves on the gamma. For instance, in the case of the sum of truncated normal variables, the shifted gamma extends the range of improvement over the normal approximation. Figure~\ref{fig:truncnorm}d shows (purple line) how this approximation almost matches that of the normal up to $\mu\sim 2\sigma$. In this case, as the third-order cumulants are also exact, the KLD gets approximated by \cite[Sec.~S1]{SM}
\begin{equation}
D_{\text{KL}}(f\|\phi_{\cG})= 
\frac{1}{48}\big(\bk_4-\bk_4^{\cG}\big)^2.
\label{eq:kappa4}
\end{equation}
Higher-order $[k/k+1]$-Pad\'e approximants would yield convolutions of shifted gammas (one for each pole of the denominator), leading to a hierarchy of empirical distributions that reflect only the aggregation of underlying positive random variables.

In summary, this approach not only provides an improvement on the CLT for positive random variables, as well as an explanation for the ubiquity of the gamma distribution, but it becomes a methodological tool to produce approximations for empirical distributions systematically more accurate.

This work has been supported by grants PID2022-140217NB-I00 (MC) and PID2022-141802NB-I00 (BASIC) (JAC), funded by MICIN/AEI/10.13039/501100011033 and by ``ERDF/EU A way of making Europe''.

\bibliography{refs}

\end{document}